\documentclass{aa}
\usepackage[varg]{txfonts} 
\usepackage{graphicx}
\usepackage{natbib,twoopt}
\usepackage[breaklinks=true,colorlinks=true,linkcolor=blue,citecolor=blue,
  urlcolor=blue]{hyperref} 
\bibpunct{(}{)}{;}{a}{}{,} 
\makeatletter
\newcommandtwoopt{\citeads}[3][][]{\href{http://adsabs.harvard.edu/abs/#3}%
{\def\hyper@linkstart##1##2{}%
\let\hyper@linkend\@empty\citealp[#1][#2]{#3}}}
\newcommandtwoopt{\citepads}[3][][]{\href{http://adsabs.harvard.edu/abs/#3}%
{\def\hyper@linkstart##1##2{}%
\let\hyper@linkend\@empty\citep[#1][#2]{#3}}}
\newcommandtwoopt{\citetads}[3][][]{\href{http://adsabs.harvard.edu/abs/#3}%
{\def\hyper@linkstart##1##2{}%
\let\hyper@linkend\@empty\citet[#1][#2]{#3}}}
\newcommandtwoopt{\citeyearads}[3][][]%
{\href{http://adsabs.harvard.edu/abs/#3}
{\def\hyper@linkstart##1##2{}%
\let\hyper@linkend\@empty\citeyear[#1][#2]{#3}}}
\makeatother

\newcommand{\heII}{He\,\textsc{ii} ($\lambda$4686)}

\begin{document}

\title
 {Exploring inside-out Doppler tomography:
  magnetic cataclysmic variables}
\author
 {E.~J.~Kotze\inst{1,2}
  \and
  S.~B.~Potter\inst{1}
  \and
  V.~A.~McBride\inst{1,2}}
\institute
 {South African Astronomical Observatory,
  PO Box 9, Observatory 7935, Cape Town, South Africa\\
  \email{ejk@saao.ac.za}\label{inst1}
  \and
  Astrophysics, Cosmology and Gravity Centre (ACGC),
  Department of Astronomy, University of Cape Town,
  Private Bag X3, Rondebosch 7701, South Africa\label{inst2}}
\date
 {Received dd Month ccyy / Accepted dd Month ccyy}
\abstract
 {Doppler tomography of magnetic cataclysmic variables is a valuable tool for
  the interpretation of the complex spectroscopic emission line profiles
  observed for these systems.}
 {We present the results of applying our inside-out velocity projection and flux
  modulation mapping techniques to the Doppler tomography of magnetic
  cataclysmic variables.}
 {The inside-out projection reverses the standard velocity direction by
  transposing the zero-velocity origin to the outer circumference and the
  maximum velocities to the origin of the velocity space.
  The inside-out tomogram is constructed by directly projecting phase-resolved
  spectra onto the inside-out framework.
  In addition, our flux modulation mapping technique extracts any information
  related to the modulation of the line flux by utilising consecutive half-phase
  tomograms.
  We apply this to both the standard and the inside-out techniques.}
 {Our test cases, the polars HU Aqr and V834 Cen, and the intermediate polar PQ
  Gem, were chosen because of their known accretion characteristics, namely
  ballistic, magnetic and curtain dominated accretion, respectively.
  In all three cases the inside-out tomogram better exposes low-velocity
  emission details which are overly compacted in the standard tomogram.
  This is especially apparent for the mid-inclination V834 Cen where the almost
  blob-like blended lower velocity emission in the standard tomogram is more
  exposed in the inside-out tomogram, making it easier to distinguish between
  the ballistic and magnetically confined accretion flows that are evident in
  the trailed spectra.
  Similarly, the inside-out tomogram enhances high velocity emission details
  which are washed out in the standard tomogram.
  This is particularly effective in revealing the high velocity magnetic
  accretion flows in the polars HU Aqr and V834 Cen.
  The addition of our flux modulation technique gives a significant improvement
  in reproducing the trailed input spectra adding more confidence to the
  interpretation of the Doppler maps.
  Furthermore, amplitude and phase maps are constructed that further reveal
  amplitude and phasing characteristics of the emission components in the three
  test cases.}
 {}
\keywords
 {accretion, accretion discs --
  techniques: spectroscopic --
  stars: binaries: close --
  stars: magnetic fields --
  stars: novae, cataclysmic variables}
\titlerunning
 {Exploring inside-out Doppler tomography (mCVs)}
\authorrunning
 {E.~J.~Kotze et al.}
\maketitle

\section{Introduction}
\label{sec:Intro}
Cataclysmic variables (CVs) are interacting, semi-detached binary systems which
consist of a white dwarf (primary) and a lower main-sequence red dwarf
(secondary).
The secondary fills its Roche lobe and loses material through the inner
Lagrangian point ($L_{1}$) of the system.
This material flows in a free fall trajectory towards the primary and it may
form an accretion disc around the primary before it is finally accreted.
In magnetic CVs (mCVs) the magnetic field of the primary is sufficiently strong
to disrupt completely or partially the formation of an accretion disc.
Instead, part of the accretion flow is threaded along magnetic field lines to
the poles of the primary in an accretion stream or an accretion curtain.
A distinction is made between two classes of mCVs, namely polars and
intermediate polars (IPs).
In polars the primary's magnetic field is sufficiently strong to lock the
primary into synchronous rotation with the binary orbit and to prevent
completely an accretion disc from forming.
In IPs the magnetic field of the primary is less strong and the primary rotates
asynchronously with the binary orbit.
The formation of a disc may be completely or partially disrupted.
\citetads[][]{2001cvs..book.....H}
and
\citetads[][]{2003cvs..book.....W}
provide comprehensive reviews of all classes of CVs and mCVs.
See also
\citetads[][]{2015SSRv..191..111F}
for an extensive review on the current observational and theoretical research of
isolated and binary magnetic white dwarfs.

Doppler tomography was introduced by
\citetads[][]{1988MNRAS.235..269M}.
It was initially developed as a technique which uses orbitally phase-resolved
spectra to construct a two-dimensional tomogram (velocity image) of the
accretion disc of a CV.
Doppler tomography has revolutionised the interpretation of complex line
profiles observed in the phase-resolved spectra of interacting binary systems
such as CVs and mCVs
\citepads[see, e.g.][]{2001LNP...573....1M}.
One of the seminal results of Doppler tomography applied to mCVs was presented
by \citetads[][]{1997A&A...319..894S}.
They were able to identify three different emission line components, with
different width and radial velocity variation, in the observed spectra of the
polar \object{HU Aqr}.
They then used Doppler tomography to locate the origin of these components (1)
on the secondary, (2) in the ballistic, and (3) in the magnetically confined
parts of the accretion flow.
Amongst other examples,
\citetads[][]{1998A&A...329..115S}
and
\citetads[][]{2004MNRAS.348..316P}
were able to do the same for \object{V2301 Oph} and \object{V834 Cen},
respectively.
\citetads[][]{1997MNRAS.288..817H, 1999ApJ...519..324H}
used spin-cycle Doppler tomography to map the spin-varying emission components
in intermediate polars.
It was especially successful in exposing the extent and shape of the
line-emitting accretion curtains in, for example, \object{PQ Gem}
\citepads[][]{1997MNRAS.288..817H, 1999ApJ...519..324H}.

Modulation Doppler tomography, a notable extension to the standard technique,
was introduced by
\citetads[][]{2003MNRAS.344..448S}.
The technique is able to isolate emission components which vary harmonically as
a function of the orbital (or spin) period.
This is achieved through the simultaneous construction of three tomograms, that
is, one for the average flux distribution and two for the variable part in terms
of its sine and cosine amplitudes.
Recent references in literature to the application of this technique mostly
involve spectroscopic studies of black hole X-ray binaries, for example,
\citetads[][]{2015MNRAS.449.1584P}
and
\citetads[][]{2015MNRAS.450.2410C}.

Inside-out Doppler tomography, a complementary extension to the standard
technique, was introduced by
\citetads[][hereafter Paper I]{2015A&A...579A..77K}.
This technique reverses the velocity axis to create an inside-out Doppler
coordinate frame.
Inside-out tomograms are constructed independently of the corresponding standard
tomograms.
This is achieved by directly projecting phase-resolved spectra onto the
inside-out framework.
The objective was to explore a more intuitive velocity layout for Doppler
tomograms that rectified their inside-out appearance.
The results presented in Paper I also showed that the inside-out projection can
enhance high velocity emission details which are washed out in the standard
projection.
Similarly, it can expose low-velocity emission details which are overly
compacted.

This paper expands on the preliminary results presented in
\citetads[][]{2016PoS...........P}.
We apply our inside-out projection technique and a variant form of modulation
Doppler tomography to the spectroscopic observations of mCVs.
First, in Sect.~\ref{sec:VelSpace} we review the polar coordinate frame for the
standard and the inside-out velocity space previously introduced.
Next, in Sect.~\ref{sec:DopTom} we do a comparison between the standard and
inside-out tomograms of three mCVs, namely the polars HU Aqr and V834 Cen, as
well as the intermediate polar PQ Gem.
We investigate how the relative contrast levels amongst emitting components such
as the ballistic and magnetically confined accretion flows are redistributed by
the inside-out framework.
In Sect.~\ref{sec:FluxMod} we present a variant form of modulation Doppler
tomography.
Our flux modulation mapping technique extracts any phased modulation in the
observed flux from consecutive half-phase standard and inside-out tomograms.
We provide a summary and conclusions in Sect.~\ref{sec:Summary}.

\section{Velocity space}
\label{sec:VelSpace}

\begin{figure}
\centering
\includegraphics[width=8.2cm]{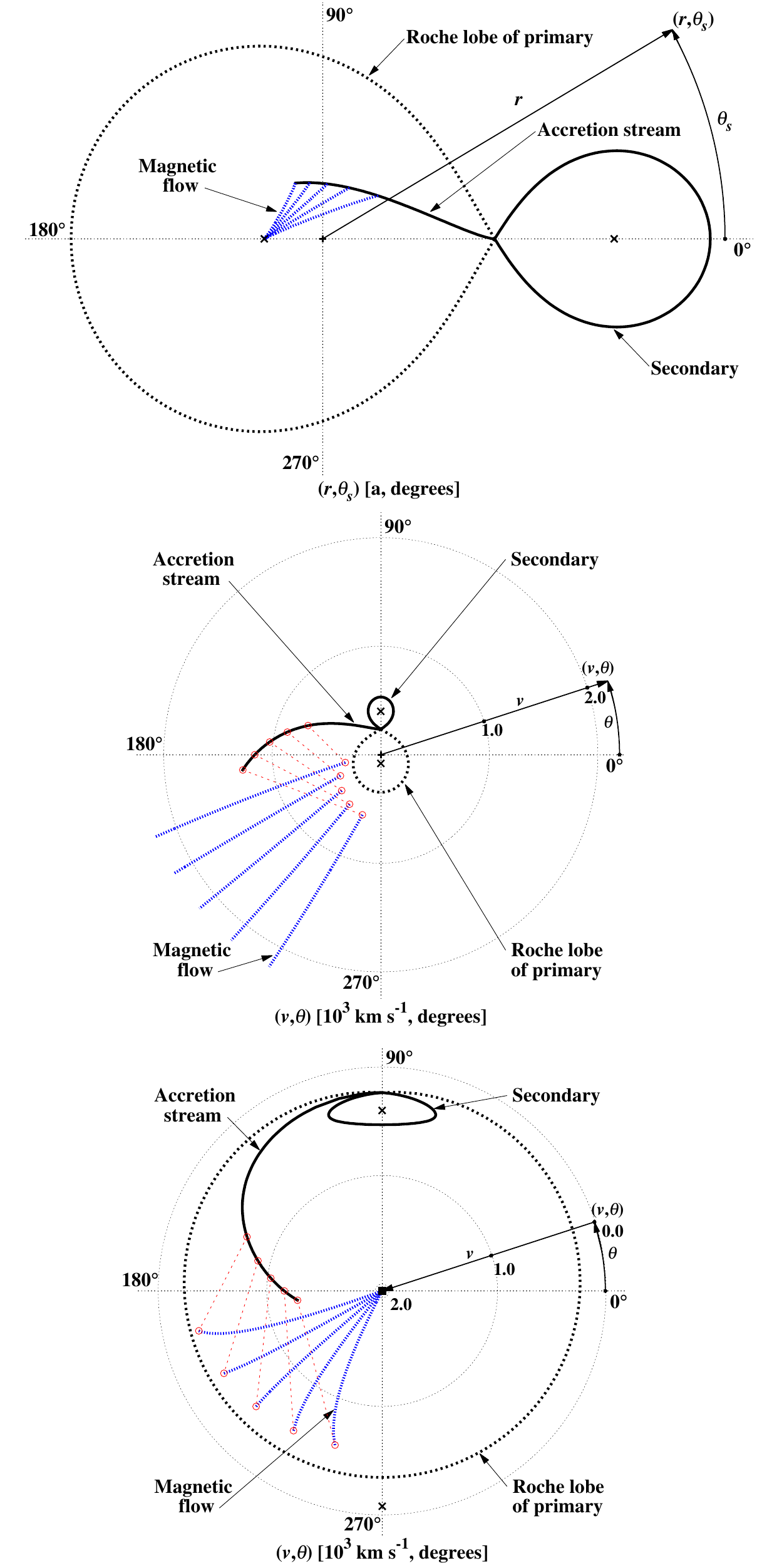}
\caption
 {
  Standard spatial and velocity, and inside-out velocity polar coordinates for
  a model mCV with ballistic and magnetic accretion flows only.
  The top and middle panels show the standard spatial and velocity polar
  coordinate layout, respectively.
  The bottom panel shows the inside-out velocity polar coordinate layout.
  The assumed binary system parameters for the model are:
  a primary mass of $0.8M_{\sun}$; a mass ratio ($q$) of $0.2$; an orbital
  period of $0.083333$ d ($\sim120$ min) and an inclination of $87\degr$.
  The C.O.M.~of the binary is marked with a plus (+) while that of the primary
  and secondary are marked with crosses ($\times$).
  Overlays are shown for the Roche lobe of the primary, the accretion stream
  from the secondary towards the primary (up to an azimuth angle of $60\degr$),
  as well as for magnetic dipole field lines at $10\degr$ intervals from
  $20\degr$ to $60\degr$ in azimuth around the primary.
  The assumed dipole axis azimuth and co-latitude are $35\degr$ and $15\degr$,
  respectively.
 }\label{fig:Polar}
\end{figure}

We first review a polar coordinate frame that co-rotates with a mCV with
ballistic and magnetic accretion flows only.
Paper I asserted that a polar coordinate frame is more conducive to the
circularly symmetric nature of Doppler tomograms.
It gives a more direct indication of velocities and directions, that is to say,
describing velocity in terms of its magnitude and direction is more intuitive
than describing it in terms of its $x$- and $y$-axial components.

The top panel in Fig.~\ref{fig:Polar} shows the spatial polar coordinates for a
model mCV without an accretion disc, but with a ballistic stream and
magnetically confined accretion along the magnetic dipole field lines.
The spatial coordinates are the radial distance $r$ from the origin, that is,
the binary's centre of mass (C.O.M.), and the polar angle $\theta_{s}$ (not to
be confused with $\theta$ in the velocity frame) measured in an anti-clockwise
direction from the line between the binary's C.O.M.~and the secondary.
The $r$ coordinate is normalised by the binary separation $a$.
The orbital motion is counter-clockwise and the point of mid-eclipse of the
primary by the secondary is defined as the binary orbital phase zero.

The middle panel in Fig.~\ref{fig:Polar} shows the corresponding velocity map in
standard polar coordinates.
The velocity magnitude $v$ increases linearly outwards from the origin and
the velocity direction is the polar angle $\theta$ measured anti-clockwise
from the line drawn from the origin horizontally to the right (i.e. the
$0\degr$-line).
For example, the secondary has a velocity magnitude of
$\sim400\mbox{\,km\,s}^{-1}$ with a velocity direction (angle) of $90\degr$.
The increasing velocity of the ballistic part of the accretion stream is
reflected in the outwards curve of the velocity profile from where it leaves
the secondary towards higher velocities.
Similarly, the profiles of the dipole trajectories radiate outwards from the
lower velocities where they leave the orbital plane towards higher velocities as
they fall down onto the primary, in other words, they diverge towards higher
velocities.

The bottom panel in Fig.~\ref{fig:Polar} shows the inside-out velocity map.
In the inside-out velocity polar coordinate frame zero velocity is on the
outer circumference and the maximum velocities are around the centre of the
coordinate frame.
In order to provide a complementary view to the standard framework the centre of
the inside-out framework is set to the maximum radial velocity extracted from
the phase-resolved spectra, that is to say, the `edge' of the data projected
onto the framework to construct the tomogram.
However, we note that the centre of the framework represents a discontinuity
and, therefore, the inner four velocity bins (pixels) around it are ignored.
The velocity magnitude $v$ now increases linearly inwards from zero velocity
towards the origin, while the velocity direction is still given by the polar
angle $\theta$ measured anti-clockwise from the $0\degr$-line.
The profile of the secondary is now upside down since it is orbiting as a
solid body with the outside moving faster than the inside.
The outer bounded circular ring area, between zero velocity and the dashed line,
is the Roche lobe of the primary.
As one can see in the inside-out framework the profile of the ballistic stream
now curves inwards as it accelerates towards higher velocities.
The profiles of the dipole trajectories now radiate inwards from the lower
velocities where they leave the orbital plane towards higher velocities as they
fall down onto the primary, in other words, they converge towards higher
velocities.
 
\section{Doppler tomography: standard and inside-out projections}
\label{sec:DopTom}

All the standard and inside-out Doppler tomograms presented hereafter were
constructed using our fast maximum entropy (FME) Doppler tomography code.
Our code is based on the FME code presented by
\citetads[][]{1998astro.ph..6141S}
and it can construct tomograms in the standard or the inside-out velocity
coordinate frame.
The inside-out tomogram is constructed independently of the standard tomogram by
directly projecting phase-resolved spectra onto the inside-out velocity space.

All the tomograms are shown with model velocity overlays.
These include the Roche lobes of the primary and the secondary as well as a
single particle ballistic trajectory from the $L_{1}$ point representing the
ballistic accretion flow.
The single particle ballistic trajectory includes gravitational and centrifugal
effects.
Also included in the ballistic trajectory is a magnetic drag force which
decreases exponentially as a function of radius.
This drag force is similar to that described by
\citetads[][]{1997A&A...319..894S}
and
\citetads[][]{2000MNRAS.313..533S},
and it is included in order to account for the pull of the primary's magnetic
field on the partly ionised stream.
The model velocity overlays also include some single particle magnetic dipole
trajectories representing the magnetically confined accretion flow.

Our test cases were selected for the distinct accretion features that are
apparent in their trailed spectra and exposed through standard Doppler
tomography.
In the standard tomogram of the eclipsing polar HU Aqr
\citepads[][]{1997A&A...319..894S}
the emission associated with a ballistic accretion stream is very prominent.
On the other hand, emission associated with a magnetically confined accretion
flow is prominent in the standard tomogram of the non-eclipsing polar V834 Cen
\citepads[][]{2004MNRAS.348..316P}.
The standard spin-cycle tomograms of the intermediate polar PQ Gem
\citepads[][]{1997MNRAS.288..817H, 1999ApJ...519..324H}
exclusively show emission associated with a magnetically confined accretion flow
locked on the spin period of the primary.

The validity and usefulness of Doppler tomograms depend on how well features in
the observed spectra are reproduced in the reconstructed spectra.
The results from our test cases show that normal Doppler tomography (standard
and inside-out projections) is able to reproduce the basic structure of the
observed spectra as well as most of the features linked to the prominent
emission components.
However, some features of the more complex structures are not fully reproduced
in all cases.
This, in part, is due to the assumption that Doppler tomography presents an
average phase map and does not take into account the flux variations phased with
the period.
In Sect.~\ref{sec:FluxMod} we show that our flux modulation mapping technique
has the ability to improve the reconstruction of the observed spectra.

For all the test cases we show normalised standard and inside-out tomograms, as
well as normalised trailed input, reconstructed and absolute
input-minus-reconstructed ($\mbox{O}-\mbox{C}$) residual spectra.
The input and reconstructed spectra are normalised by the maximum flux
in the input spectra.
We choose to show absolute $\mbox{O}-\mbox{C}$ spectra in order to have all the
displayed trailed spectra on a comparable colour scale and calculate a simple
root mean square (rms) value for the residuals, that is,
\begin{equation}
 \mbox{rms}=\sqrt{\frac{1}{n}\sum_{i=1}^{n}\left(\mbox{O}_{i}-\mbox{C}_{i}\right)^{2}},
\end{equation}
where $n$ is the total number of data points.
All the tomograms, standard and inside-out, were constructed using $3996$
discrete velocity bins (pixels).
We choose a slightly higher value for the regularisation parameter $\alpha$,
that is, $0.003$ as opposed to the default value of $\sim0.002$, as described by
\citetads[][]{1998astro.ph..6141S}.
This was used for all the tomograms so that all are on the same comparable
measure of regularisation.
Also, it produces solutions that reproduce the detail in the observed spectra,
but it does not over-fit the noise.

\subsection{The polar HU Aqr}
\label{sec:DTHUAqr}

\begin{figure*}
\centering
\includegraphics[width=16cm]{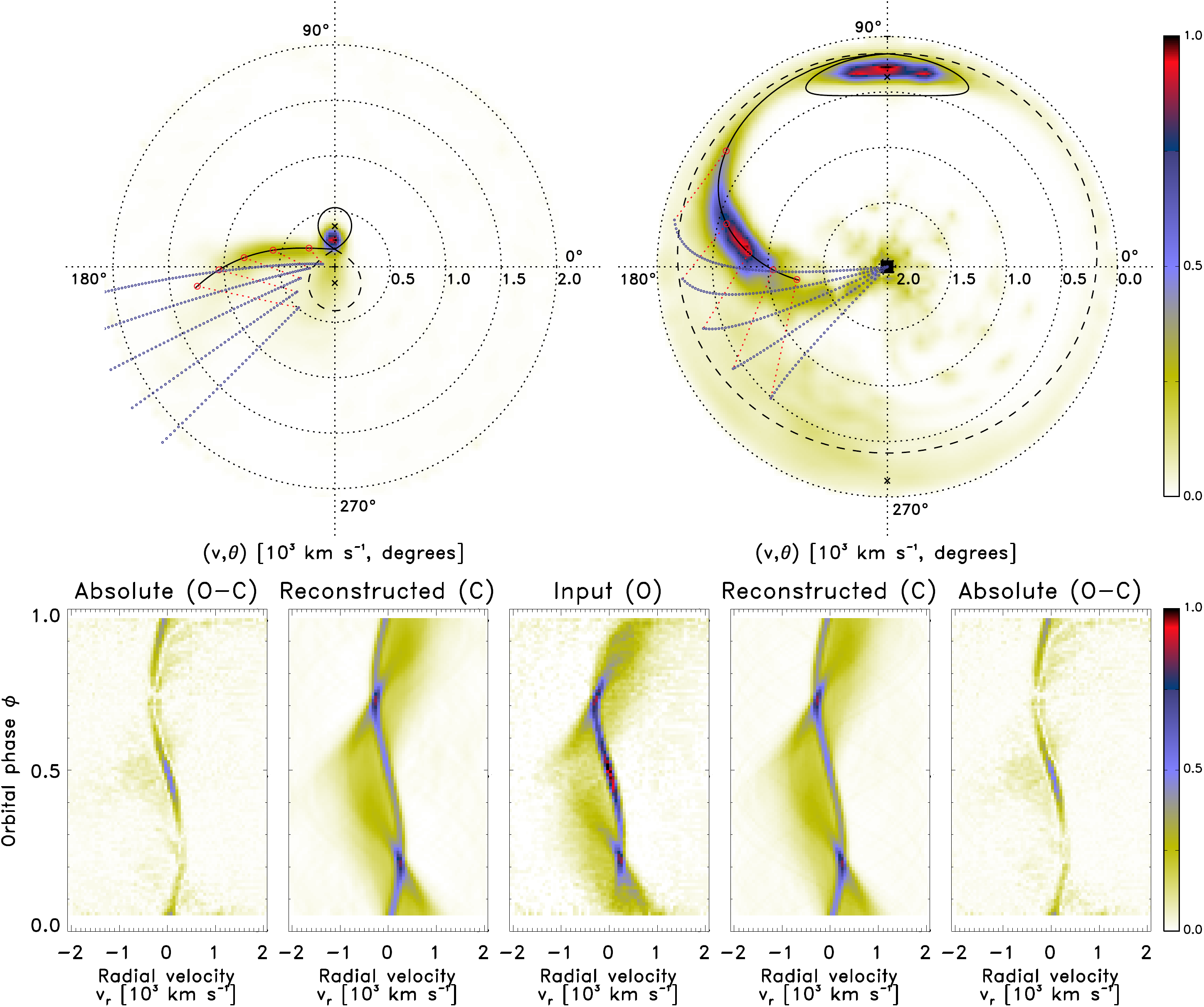}
\caption
 {
  Doppler tomography of HU Aqr.
  The standard and inside-out tomograms are shown top left and right,
  respectively.
  The bottom middle panel shows the trailed input spectra.
  The panels to the left and right of the middle panel show the trailed
  reconstructed and absolute residual spectra for the corresponding tomogram.
  The model velocity profile overlay shown in both tomograms is based on one of
  the models used by
  \citetads[][]{1997A&A...319..894S}
  with an inclination angle of $84\degr$, a primary mass of $0.875M_{\sun}$, a
  mass ratio $q=0.40$ and an orbital period of $0.086820446$ d ($\sim125$ min).
  The overlay includes the Roche lobes of the primary (dashed line) and the
  secondary (solid line) as well as a single particle ballistic trajectory from
  the $L_{1}$ point up to $45\degr$ in azimuth around the primary (solid line).
  Magnetic dipole trajectories are calculated at $10\degr$ intervals from
  $5\degr$ to $45\degr$ in azimuth around the primary (thin dotted lines).
  The first dipole trajectory starts at ($123\mbox{\,km\,s}^{-1}$, $167\degr$),
  with consecutive trajectories starting at locations with progressively higher
  velocities and polar angles.
  The first dipole connection (small circles) is at ($340\mbox{\,km\,s}^{-1}$,
  $145\degr$).
  The dipolar axis azimuth and co-latitude, as modelled by
  \citetads[][]{1999MNRAS.304..145H},
  are $\sim38\degr$ and $\sim12\degr$, respectively.
 }\label{fig:dtHUAqr}
\end{figure*}

HU Aqr is a bright eclipsing polar.
Standard Doppler tomography of HU Aqr presented by
\citetads[][]{1997A&A...319..894S}
showed an extraordinary prominent ballistic stream, at least compared to other
polars, distinct emission from the secondary and a diffuse patch of emission
thought to be associated with the magnetically confined accretion flow.
HU Aqr is therefore an excellent candidate as a test case for inside-out
Doppler tomography.
Figure \ref{fig:dtHUAqr} shows the results of applying standard and inside-out
Doppler tomography to the \heII~emission line from 1993 August spectroscopic
observations of the high accretion state of HU Aqr.
These observations were presented first by
\citetads[][]{1997A&A...319..894S}
and the reader is referred to that paper for a description of the data set and
reduction procedures.
Because HU Aqr is an eclipsing system, we exclude a small range around phase
$0.0$ and only use the $92$ spectra covering the orbital phase range
$0.06 - 0.97$.
The spectra are evenly spaced and each spectrum has $72$ velocity points in the
extracted radial velocity range around the \heII~emission line.

The trailed input spectra show three distinct emission components.
The first component is the prominent narrow line which is brightest around phase
$0.5$ and which has zero radial velocity at phase $0.0$.
This component has a low-velocity amplitude and is associated with emission from
the irradiated secondary.
The second component is also relatively narrow, but it has a high velocity
amplitude.
It has a maximum redshift around phase $0.95$ and a maximum blueshift around
phase $0.45$.
This component crosses the first component around phases $0.2$ and $0.7$.
The third component is relatively broad and visible throughout the covered phase
range.
It has a maximum blueshift around phase $0.40$.
Both the second and third component are associated with emission produced in
different parts of the accretion flow.

The emission from the irradiated secondary is seen as a bright spot at
($360\mbox{\,km\,s}^{-1}$, $90\degr$) in both the standard and inside-out
tomograms.
This feature in the tomograms is well traced by the velocity profile of the
Roche lobe of the secondary.
In the standard tomogram the secondary dominates the brightness scale due to its
compacted projection.
In contrast, its more extended projection in the inside-out tomogram allows
other emission features to appear brighter.

In the standard tomogram the emission associated with the ballistic part of the
accretion stream is a prominent, slightly downward-sloping horizontal ridge with
an apparent relative constant brightness.
However, a closer inspection reveals that at first, as the stream leaves $L_{1}$
at ($200\mbox{\,km\,s}^{-1}$, $90\degr$), it is faint, but it becomes brighter
as it reaches ($300\mbox{\,km\,s}^{-1}$, $135\degr$).
It keeps a relative consistent brightness up to
($750\mbox{\,km\,s}^{-1}$, $170\degr$) before it starts to become fainter.
It stays clearly discernible up to ($1000\mbox{\,km\,s}^{-1}$, $180\degr$).
After this point it becomes more diffused and almost no detail is discernible
at velocities above $1000\mbox{\,km\,s}^{-1}$.
This distribution in brightness along the ballistic stream is more prominent in
the inside-out tomogram, that is, the lower velocity part is even fainter,
whereas the mid- and high-velocity parts are significantly brighter.
The lower velocity emission appears fainter as the emission is re-binned in a
larger area with more pixels.
Conversely, the high velocity emission is re-binned in a smaller area with fewer
pixels.
The mid-velocity emission, especially in the range
($500-1000\mbox{\,km\,s}^{-1}$, $160-180\degr$), appears brighter because
overall the relative contrast levels changed due to the more extended projection
of the secondary.
In both tomograms the brighter part of the ballistic stream is well traced by
the modelled velocity profile for the threading region, that is, the region
where the ballistic stream is threaded onto the magnetic field lines, but before
it leaves the orbital plane.
Modelling done by 
\citetads[][]{1999MNRAS.304..145H}
on the magnetic stripping of the ballistic stream showed that it is possible for
the stream to `survive' up to a distance of $0.5a$ along its trajectory.
This is equivalent to a velocity of $\sim1250\mbox{\,km\,s}^{-1}$.
It is therefore possible that part of the emission visible in the inside-out
tomogram along the extended ballistic trajectory in the velocity range
($1000-1500\mbox{\,km\,s}^{-1}$, $180-220\degr$) can be ascribed to the
ballistic stream.
This tenuous higher velocity ($>1000\mbox{\,km\,s}^{-1}$) emission is not
clearly discernible in the standard tomogram.
Resolving it in the inside-out tomogram is possible as the emission is re-binned
in a smaller area with fewer pixels.

The diffuse patch of emission seen in the standard tomogram in the velocity
range ($0-500\mbox{\,km\,s}^{-1}$, $180-270\degr$) is considered to be
associated with the magnetically confined accretion flow that has left the
orbital plane.
In the inside-out tomogram this patch of emission is exposed to extend to even
higher velocities, covering a velocity range of ($0-1000\mbox{\,km\,s}^{-1}$,
$180-270\degr$).

The three distinct line components identified in the input spectra are
reproduced fairly well in the reconstructed spectra from both the standard and
inside-out tomograms.
However, neither of the projections is able to reproduce the observed flux
distribution in the prominent narrow line component associated with emission
from the irradiated secondary.
This is clearly seen in both sets of absolute $\mbox{O}-\mbox{C}$ spectra.
Overall, the two projections performed equally well in reproducing the input
spectra, that is, the residuals of both have rms values of $0.043$ ($n = 6624$).
In Sect.~\ref{sec:FMHUAqr} we show the improvement in the reproduction of the
narrow feature achieved with our flux modulation mapping technique.

\subsection{The polar V834 Cen}
\label{sec:DTV834Cen}

\begin{figure*}
\centering
\includegraphics[width=16cm]{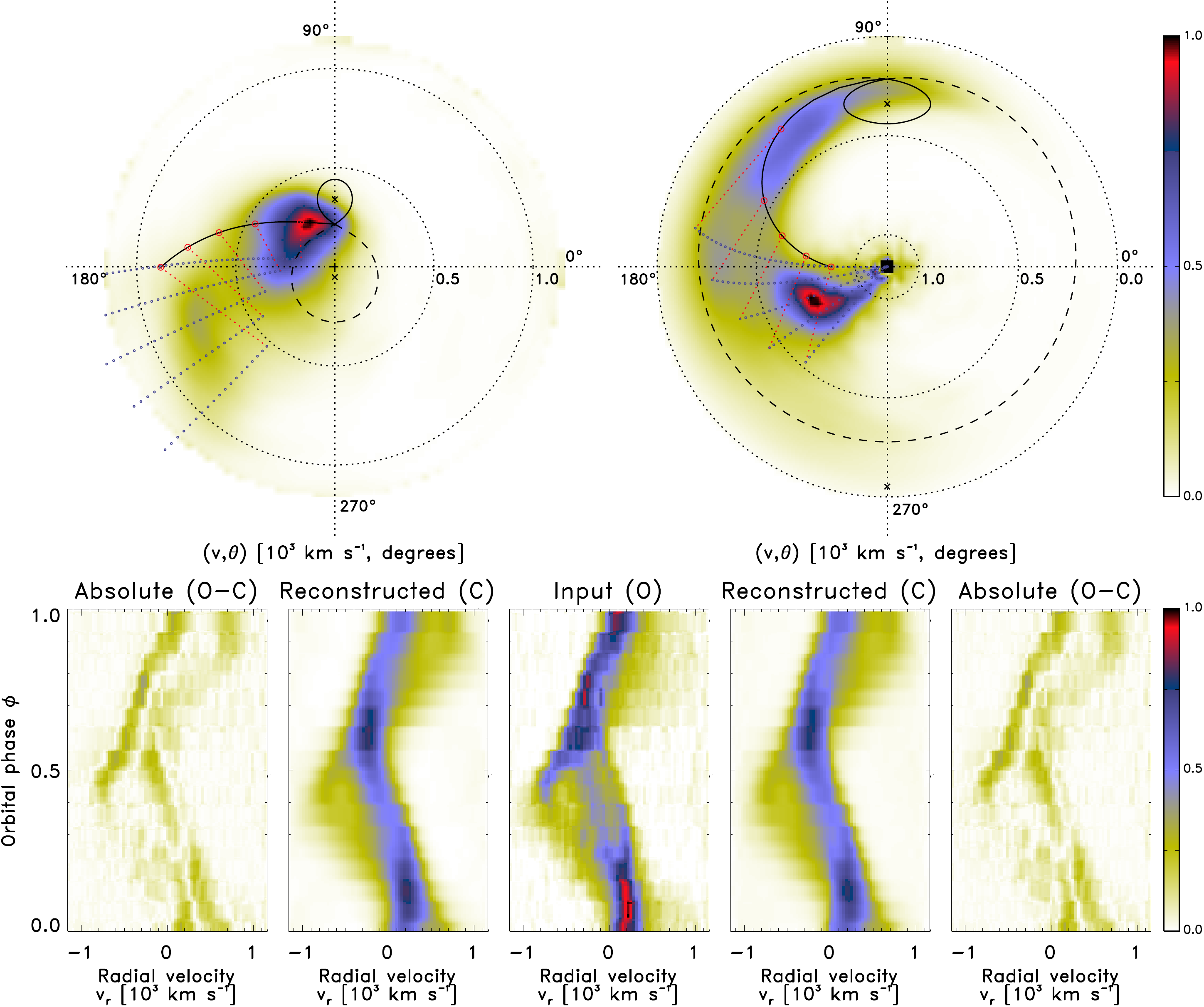}
\caption
 {
  Doppler tomography of V834 Cen.
  Same layout as Fig.~\ref{fig:dtHUAqr}.
  The model velocity profile overlay shown in both tomograms is based on the
  model used by
  \citetads[][]{2004MNRAS.348..316P}
  with an inclination angle of $50\degr$, a primary mass of $0.85M_{\sun}$, a
  mass ratio $q=0.154$ and an orbital period of $0.070497518$ d
  ($\sim101.5$ min;
  \citeads[][]{1993A&A...267..103S}).
  The Roche lobes of the primary (dashed line) and the secondary (solid line)
  as well as a single particle ballistic trajectory from the $L_{1}$ point up
  to $45\degr$ in azimuth around the primary (solid line), are included in the
  overlay.
  Magnetic dipole trajectories are calculated from $5\degr$ to $45\degr$ in
  azimuth around the primary (thin dotted lines) at $10\degr$ intervals using a
  dipolar axis azimuth and co-latitude of $\sim36\degr$ and $\sim20\degr$
  \citepads[][]{2004MNRAS.348..316P},
  respectively.
  The first of the dipole connections (small circles) are at
  ($360\mbox{\,km\,s}^{-1}$, $135\degr$) and the first dipole trajectory starts
  at ($150\mbox{\,km\,s}^{-1}$, $162\degr$).
  Consecutive trajectories start at locations with progressively higher
  velocities and polar angles.
 }\label{fig:dtV834Cen}
\end{figure*}

V834 Cen is a bright non-eclipsing polar.
\citetads[][]{2004MNRAS.348..316P}
presented standard Doppler tomography of V834 Cen that exposed dominant emission
from the secondary and (or) from around $L_{1}$, less discernible emission from
the ballistic stream and extended diffuse emission considered to be consistent
with the magnetically confined accretion flow.
Figure \ref{fig:dtV834Cen} shows standard and inside-out Doppler tomography
based on the \heII~emission line from the 2000 August spectroscopic observations
of V834 Cen presented by
\citetads[][]{2004MNRAS.348..316P}.
The reader is referred to that paper for more information on the data set and
reduction procedures.
We use all $59$ spectra from the original data set, which covers the whole
orbital phase range, for our calculations.
Each spectrum covers $0.05$ of the phase with some overlap and has $73$ velocity
points in the extracted radial velocity range around the \heII~emission line.

The trailed input spectra have a complex structure with at least three emission
components.
There is a blended low- to mid-velocity component which is brightest between
phases $0.0-0.2$.
It is fainter, but separated between phases $0.2-0.55$, and again blended and
brighter between phases $0.55-1.0$.
This component is associated with blended emission from the irradiated secondary
and the low- to mid-velocity part of the ballistic accretion stream.
Also discernible is a fainter high-velocity broad base component underlying the
whole phase range.
The broad base component is associated with emission produced in different parts
of the accretion flow.
There is also a striking brighter blueshifted wing that follows closely the
high-velocity edge of the underlying broad base component between phases
$0.4-0.55$.
This high-velocity component is associated specifically with emission produced
in the magnetically confined accretion flow as it falls towards the primary.

In the standard tomogram the emission from the irradiated secondary is blended
with emission from the ballistic stream.
This overly compacted blended emission creates an extremely bright ridge in the
velocity range ($250-500\mbox{\,km\,s}^{-1}$, $90-150\degr$) and dominates the
brightness scale of the standard tomogram.
In the inside-out tomogram this blended emission is more separated:
The secondary is seen as a diffused patch at ($340\mbox{\,km\,s}^{-1}$,
$90\degr$),
while the stream forms a brighter ridge along the model stream trajectory in the
velocity range ($300-400\mbox{\,km\,s}^{-1}$, $115-140\degr$).
Separating this lower velocity emission in the inside-out tomogram is possible
as the emission is re-binned in a larger area with more pixels.

The emission associated with the magnetically confined accretion as it leaves
the orbital plane in the threading region is seen in both tomograms in the
velocity range ($0-500\mbox{\,km\,s}^{-1}$, $170-230\degr$).
In the inside-out tomogram, however, the separation between the lower velocity
emission from the ballistic stream and that from the magnetically confined
accretion is more pronounced.
Also, in both tomograms the emission considered to be consistent with the
magnetically confined accretion as it flows down to the primary is centred on
the ridge of brightness covering the velocity range
($700-1000\mbox{\,km\,s}^{-1}$, $185-235\degr$).
However, in the inside-out tomogram the funnelling of the emission as it falls
towards the primary along the dipole trajectories, especially those with
azimuth angles $35\degr$ and $45\degr$, is significantly more apparent than in
the standard tomogram.

The basic structure of the emission components seen in the input spectra is
reproduced in the reconstructed spectra from both the standard and inside-out
tomograms.
However, the more complex structure between phases $0.2-0.55$ and the observed
flux distribution in the narrower component between phases $0.0-0.2$ and
$0.55-1.0$ are not reproduced by either of the projections.
The absolute $\mbox{O}-\mbox{C}$ spectra of both projections clearly show the
incomplete reconstructed parts.
Both projections, however, achieved the same level of overall success, that is,
the residuals of both have rms values of $0.076$ ($n = 4307$).
We show in Sect.~\ref{sec:FMV834Cen} that our flux modulation mapping
technique is able to reproduce the complex structure and the flux distribution.

\subsection{The intermediate polar PQ Gem}
\label{sec:DTPQGem}

\begin{figure*}
\centering
\includegraphics[width=16cm]{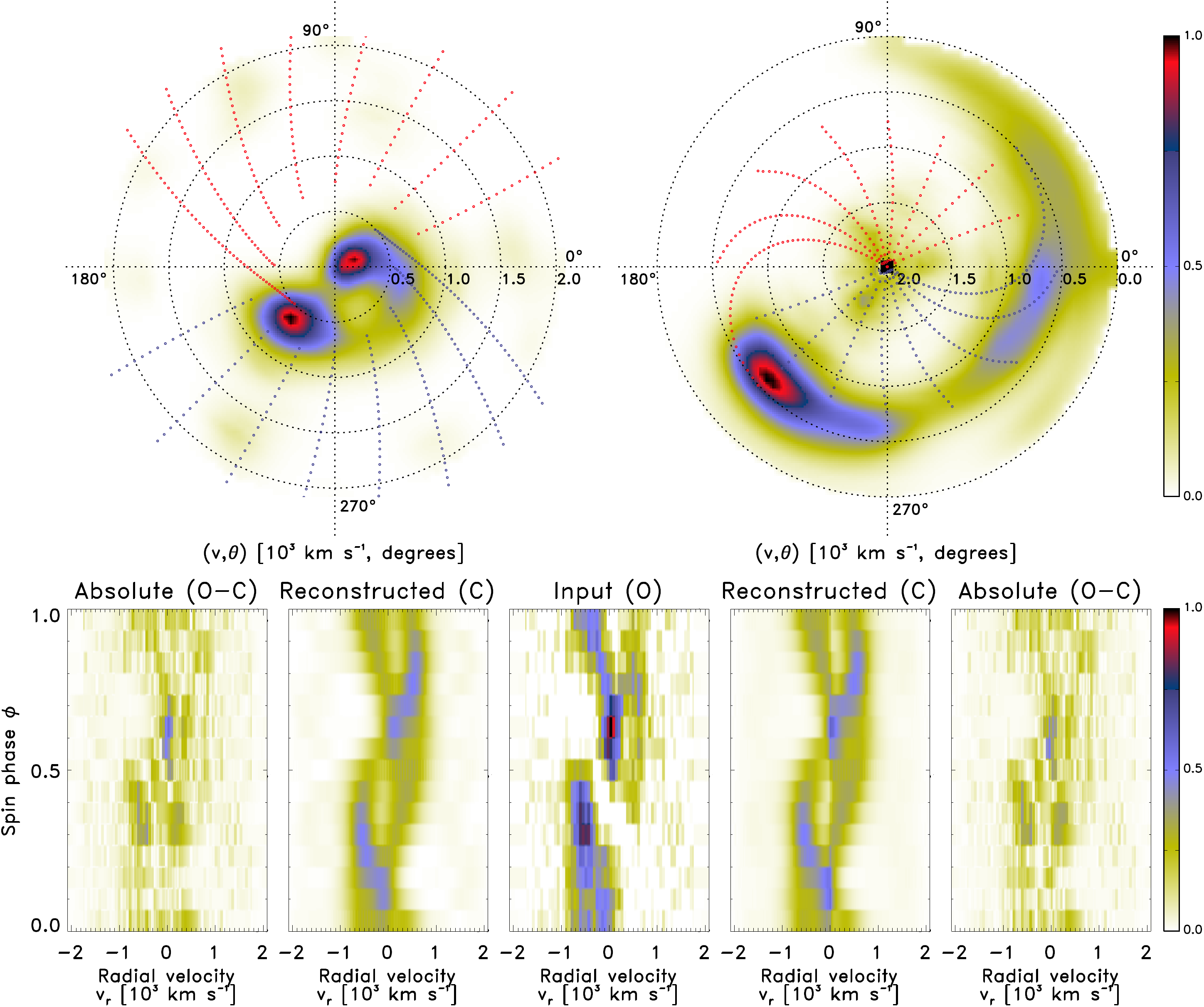}
\caption
 {
  Spin-cycle Doppler tomography of PQ Gem.
  The standard and inside-out spin-cycle tomograms are shown top left and right,
  respectively.
  The bottom middle panel shows the trailed input spectra.
  The panels to the left and right of the middle panel show the trailed
  reconstructed and absolute residual spectra for the corresponding spin-cycle
  tomogram.
  The model velocity profile overlay shown in both tomograms is based on an
  inclination angle of $30\degr$
  \citepads[][]{1997MNRAS.288..817H}
  and a dipolar axis co-latitude of $30\degr$
  \citepads[][]{1997MNRAS.285...82P}.
  As proposed by
  \citetads[][]{1997MNRAS.288..817H}
  we use a spin-wave velocity amplitude of $\sim500\mbox{\,km\,s}^{-1}$ and let
  the upper magnetic pole point towards us at spin phase $0.87$ (dipolar axis
  azimuth $\sim 47\degr$).
  We allow the upper curtain (thin blue dotted lines) to extend over spin phase
  $0.63-0.96$ and the lower curtain (thin red dotted lines) $0.13-0.46$.
 }\label{fig:dtPQGem}
\end{figure*}

PQ Gem is a bright, low-inclination ($i\sim30\degr$;
\citeads{1997MNRAS.288..817H})
intermediate polar.
\citetads[][]{1997MNRAS.288..817H, 1999ApJ...519..324H}
presented spin-cycle Doppler tomography of PQ Gem (see the mentioned papers for
a complete description of the technique).
Spin-cycle tomography is uniquely geared to produce velocity maps of the
emission components that are locked on the spin period of the primary, for 
example the accretion curtains of the magnetically confined accretion flow
falling onto the primary.
Figure \ref{fig:dtPQGem} shows standard and inside-out spin-cycle Doppler
tomography based on the spin-folded \heII~emission line from the 1993 December
spectroscopic observations of PQ Gem presented first by
\citetads[][]{1997MNRAS.288..817H}.
The reader is referred to that paper for a detailed description of the data set
and reduction procedures.
All $15$ spectra from the original spin-folded data set, which covers the whole
spin phase range, are used for our calculations.
The spectra are evenly spaced and each spectrum has $361$ velocity points in the
extracted radial velocity range around the \heII~emission line.

At least two distinct emission components are seen in the spin-folded trailed
input spectra.
The first component has a maximum blueshift around phase $0.4$ and a maximum
redshift around phase $0.9$.
It is bright at phases $0.3$ and $0.65$, but the expected blue-to-red crossover
is not clear.
The second component has a lower velocity amplitude with a maximum redshift
around phase $0.4$ and a maximum blueshift around phase $0.9$.
Although the red-to-blue crossover is visible for this component, the
blue-to-red crossover is again not clear.
Both the components are associated with emission produced in the spin-locked
magnetically confined accretion flow.

The standard tomogram of PQ Gem is dominated by two bright spots, one in the
upper right quadrant
and the other in the lower left quadrant.
\citetads[][]{1997MNRAS.288..817H, 1999ApJ...519..324H}
suggested that these two spots are associated with emission from the accretion
curtains flowing towards, respectively, the lower and upper magnetic poles of
the primary.
The accretion curtains are considered to be formed by material at the
magnetosphere flowing onto the magnetic field lines at an azimuth $30-50\degr$
ahead of the accretion region on the primary
\citepads[][]{1997MNRAS.288..817H}.
In the inside-out tomogram the spot seen in the upper right quadrant of the
standard tomogram, is projected as an extended, more diffused feature.
The brighter of the two spots seen in the standard tomogram, in other words, the
one in the lower left quadrant, is also more extended, but still the most
prominent feature.

The inside-out projection significantly changes the apparent flux distribution
between these two emission features.
This places greater emphasis on the difference in the velocities of the two
features.
If we consider the model velocity profile we find that in the inside-out
projection the magnetic dipole lines converge towards higher velocities.
This creates an intuitive representation of the curving accretion curtains as
they fall down onto the primary.
We note that the phase resolution (i.e. only 15 phase bins) of the data for PQ
Gem may be considered inadequate to allow for any significant new insights.

The basic structure of the components identified in the input spectra is
reproduced in the reconstructed spectra from both the standard and inside-out
tomograms.
However, the observed flux distribution, especially between phases $0.5-1.0$, is
not reproduced by either of the projections. 
This partial reproduction is evident in the excess seen in both sets of absolute
$\mbox{O}-\mbox{C}$ spectra.
The inside-out projection achieved a marginal better overall result, that is,
its residuals have a rms value of $0.092$ compared to the $0.094$ of the
standard projection ($n = 5415$).
In Sect.~\ref{sec:FMHUAqr} we show how our flux modulation mapping technique
improves the reproduction of the flux distribution.

\section{Doppler tomography: flux modulation mapping}
\label{sec:FluxMod}

One of the axioms of Doppler tomography
\citepads[][]{2001LNP...573....1M}
is that the technique assumes that all points in the binary system being mapped
are equally visible at all times.
This implies that it is possible to construct a tomogram using spectra
covering half of a phase only.
The fact that this axiom is violated in most CVs and mCVs (especially in the
higher inclination systems) actually enables us to selectively eliminate various
emission components from the tomogram by using spectra from half of a phase
only, thus revealing less obvious components
\citepads[e.g.][]{2004MNRAS.348..316P}.

Another axiom of Doppler tomography
\citepads[][]{2001LNP...573....1M}
is that the observed flux from any point in the binary system is constant.
Observations of CVs and mCVs, however, confirm that the flux from the typical
emission components being mapped does modulate.
Phased flux modulations are mostly attributed to the geometry of the system,
for example, an eclipse or the aspect variation of an emission component.
Doppler tomography, however, presents a phase-averaged map of the emission
distribution in the system.
Phase-dependent details in the observed spectra such as the orbital flux
modulation will therefore not be recovered in the reconstructed spectra.
Recognising this,
\citetads[][]{2003MNRAS.344..448S}
introduced modulation Doppler tomography, a technique that maps emission
components which modulate sinusoidally as a function of the orbital (or spin)
period.
This is achieved through the simultaneous construction of three tomograms, that
is, one for the average flux distribution and two for the variable part in terms
of its sine and cosine amplitudes.

Exploiting the principles introduced by 
\citetads[][]{2003MNRAS.344..448S}
and
\citetads[][]{2004MNRAS.348..316P}
we present a variant form of modulation Doppler tomography.
Our flux modulation mapping technique produces Doppler maps that represent the
amplitude, phase and average of the modulated emission.
By extracting any phased modulation in the observed flux from a series of
consecutive half-phase tomograms it is possible to map how the flux from a
specific emission component modulates over a complete phase.
If it is assumed that the flux from an emission component varies harmonically
over the observed phases then it is possible to extract the amplitude ($A_{j}$),
phase-offset ($\varphi_{j}$) and average ($B_{j}$) of the flux modulation in
the $j$-th pixel from a simple sinusoidal fit of the form
\begin{eqnarray}
\label{eq:sin_fit}
F_{ij}=A_{j}\sin\left[2\pi\left(\Phi_{i}-\varphi_{j}\right)\right]+B_{j}.
\end{eqnarray}
$F_{ij}$ and $\Phi_{i}$ are, respectively, the flux in the $j$-th pixel and the
mid-phase value of the $i$-th half-phase tomogram.
The fitted amplitude for each pixel ($A_{j}$) produces an amplitude map which
helps to identify which emission components are modulated.
The normalised amplitude map is presented with the same colour scheme as the
normal tomogram.
Adding $0.25$ to the fitted phase-offset for each pixel ($\varphi_{j}$) produces
a map for the phase of maximum flux which shows at which phase an emission
component appears brightest to an observer.
This map is colour-coded according to phase: $0.0$ - black, $0.25$ - red,
$0.5$ - green and $0.75$ - blue.
The average for each pixel ($B_{j}$) is effectively the normal tomogram in the
standard and inside-out projections.
The flux modulation mappings presented hereafter are based on $10$ consecutive
half-phases (i.e. $0.0-0.5$, ..., $0.4-0.9$, $0.5-0.0$, $0.6-0.1$, ...,
$0.9-0.4$).
We also show normalised trailed summed input and reconstructed, as well as
absolute $\mbox{O}-\mbox{C}$ spectra for all the test cases in a similar way as
in the previous section.

\subsection{The polar HU Aqr}
\label{sec:FMHUAqr}

\begin{figure*}
\centering
\includegraphics[width=16cm]{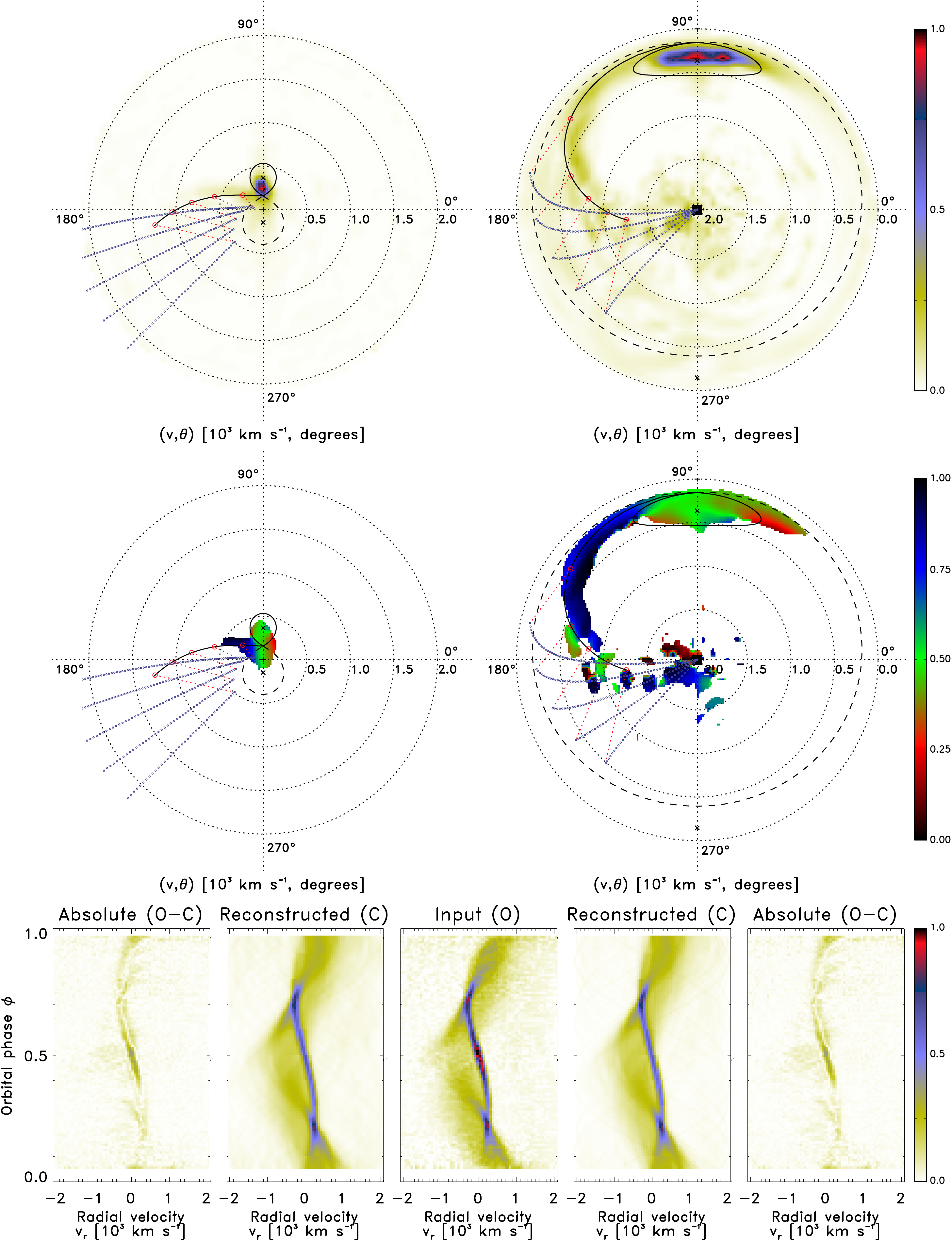}
\caption
 {
  Flux modulation mapping of HU Aqr.
  The standard and inside-out velocity maps are shown left and right,
  respectively.
  The top two panels show the modulation amplitude maps and the middle two
  panels show the phase of maximum flux maps.
  The phase of maximum flux map shows only pixels where the corresponding
  modulation amplitude is at least $7.5\%$ of the maximum amplitude.
  This map is colour coded to represent phase: black (0.0), red (0.25), green
  (0.5) and blue (0.75).
  The bottom middle panel shows the trailed input spectra.
  The panels to the left and right of the middle panel show the summed trailed
  reconstructed and absolute residual spectra for the corresponding
  $10$ consecutive half-phase tomogram.
  The model velocity profile overlay shown in all the tomograms is the same as
  the one in Fig.~\ref{fig:dtHUAqr}.
 }\label{fig:fmHUAqr}
\end{figure*}
Figure \ref{fig:fmHUAqr} shows standard and inside-out flux modulation mapping
of HU Aqr.
When compared with the reconstructed spectra in Fig.~\ref{fig:dtHUAqr} the
summed reconstructed spectra better reproduce the input spectra.
Most striking is the almost complete absence of the bright narrow line flux
between phases $0.0-0.2$ and $0.8-1.0$.
This narrow line is associated with emission from the irradiated side of the
secondary.
It is expected to be absent in the range around phase $0.0$ when the
non-irradiated side of the secondary is pointing towards us as it eclipses the
primary.
The brightening around phase $0.5$ in the same bright narrow emission is
also slightly better reproduced.
The improvement achieved for both projections is clearly seen in their absolute
$\mbox{O}-\mbox{C}$ spectra with rms values of $0.026$ compared to the
non-modulated rms values of $0.043$.

In both the standard and inside-out amplitude maps we see that the secondary is
the most flux modulated component.
The flux from the ballistic stream is also shown to modulate, but more so in the
inside-out map.
HU Aqr is a high-inclination eclipsing system and we can therefore expect that
the flux from the irradiated side of the secondary and the ballistic stream will
modulate over the orbital phase of the system as these components get eclipsed.

In the phase of maximum flux maps the secondary is mostly green which indicates
that it appears brightest around orbital phase $0.5$, when the irradiated side
of the secondary is pointing towards us.
The trailing and leading sides of the secondary are red and blue, respectively.
This is expected since we have full views of the trailing and leading sides at
orbital phases $0.5$ and $0.75$, respectively.
The ballistic stream is a mixture of blue and black which means it appears
brightest between orbital phases $0.75$ and $0.0$, in other words, between when
we have a full view of the leading side of the stream and when the stream starts
flowing away from us.
The standard amplitude map and the phase of maximum flux map hold amplitude and
phase information for the ballistic stream up to velocities of $\sim500$
km s$^{-1}$ whereas the inside-out versions expose information up to velocities
of $\sim1000\mbox{\,km\,s}^{-1}$.

\subsection{The polar V834 Cen}
\label{sec:FMV834Cen}

\begin{figure*}
\centering
\includegraphics[width=16cm]{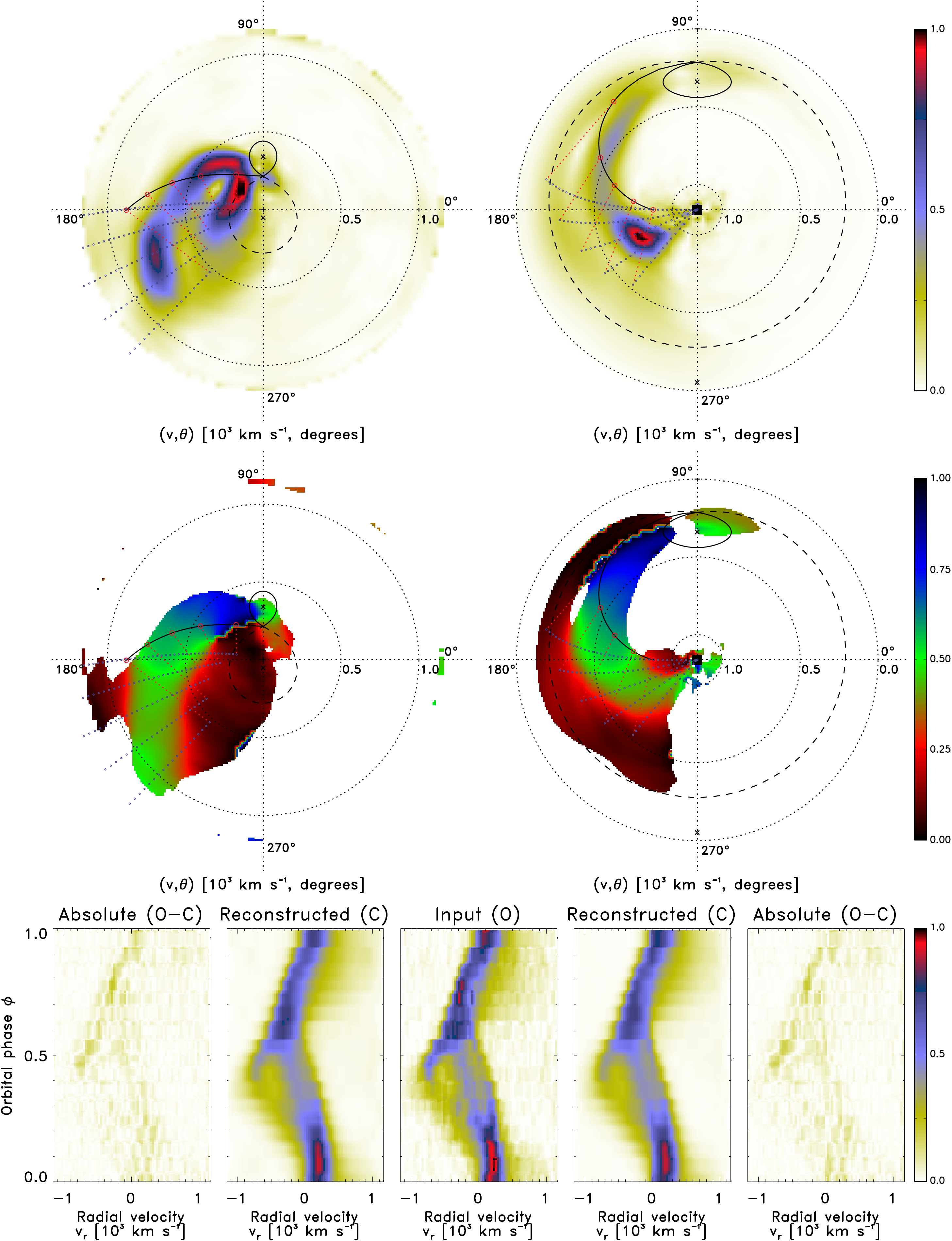}
\caption
 {
  Flux modulation mapping of V834 Cen.
  Same layout as Fig.~\ref{fig:fmHUAqr}.
  The model velocity profile overlay shown in all the tomograms is the same as
  the one in Fig.~\ref{fig:dtV834Cen}.
 }\label{fig:fmV834Cen}
\end{figure*}
The standard and inside-out flux modulation mapping results for V834 Cen are 
shown in Fig.~\ref{fig:fmV834Cen}.
The summed reconstructed spectra better reproduce the input spectra when
compared with the reconstructed spectra in Fig.~\ref{fig:dtV834Cen}.
Most prominent is the considerable brightening in the narrower line flux between
phases $0.0-0.2$ and $0.55-1.0$.
Also, the more complex, multiple flux components between phases $0.2-0.55$, and
the flux level of the striking high-velocity blue wing at phase $0.4-0.55$ are
better reproduced.
Both sets of absolute $\mbox{O}-\mbox{C}$ spectra clearly show the significant
improvement achieved with rms values of $0.030$ and $0.035$ compared to the
non-modulated rms values of $0.076$.

The most prominent flux modulated components in both the standard and inside-out
amplitude maps are the ballistic and the magnetically confined accretion flows.
In the inside-out amplitude map it almost looks like the ballistic stream is
`split' into two parts, that is, one with velocities less than and the other
with velocities greater than $250\mbox{\,km\,s}^{-1}$.
However, this is a neat separation of the magnetically confined accretion flow
as it leaves the orbital plane along the early part of the ballistic stream
(velocities $<250\mbox{\,km\,s}^{-1}$) and the mid-velocity part of the `real'
ballistic stream (velocities $>250\mbox{\,km\,s}^{-1}$).
The secondary is also flux modulated, but since V834 Cen is a mid-inclination
non-eclipsing system the amplitude of the modulation is not as dominant as that
of the secondary of HU Aqr, which is an eclipsing system.

The real ballistic stream starts out as black in both the phase of maximum flux
maps, then it switches to blue and a mixture of blue and green.
This means that the lower velocity part of the stream appears brightest at
orbital phase $0.0$ when it is pointing away from us, while the mid-velocity
part appears brightest at orbital phase $0.75$ when we have a full view of the
outer side of the stream.
The phase of maximum flux interpretation for the higher velocity part of the
ballistic stream is ambiguous since emission from this part is blended with
emission from the magnetically confined accretion flow.
The mid- to high-velocity part of the threading region is predominantly red in
both maps, that is to say, it appears brightest at orbital phase $0.25$ when we
have a full view of the inner sides of the ballistic and magnetically confined
accretion flows.
The low-velocity part of the threading region, on the other hand, is black and
red, which means that it appears brightest between orbital phases $0.0$ and
$0.25$ when it is predominantly flowing away from us.
The higher velocity part of the magnetically confined accretion flow is mostly
green, in other words, it appears brightest around orbital phase $0.5$ when we
have a full view of the outer side of the magnetically confined accretion flow
with most of it flowing towards us.

\subsection{The intermediate polar PQ Gem}
\label{sec:FMPQGem}

\begin{figure*}
\centering
\includegraphics[width=16cm]{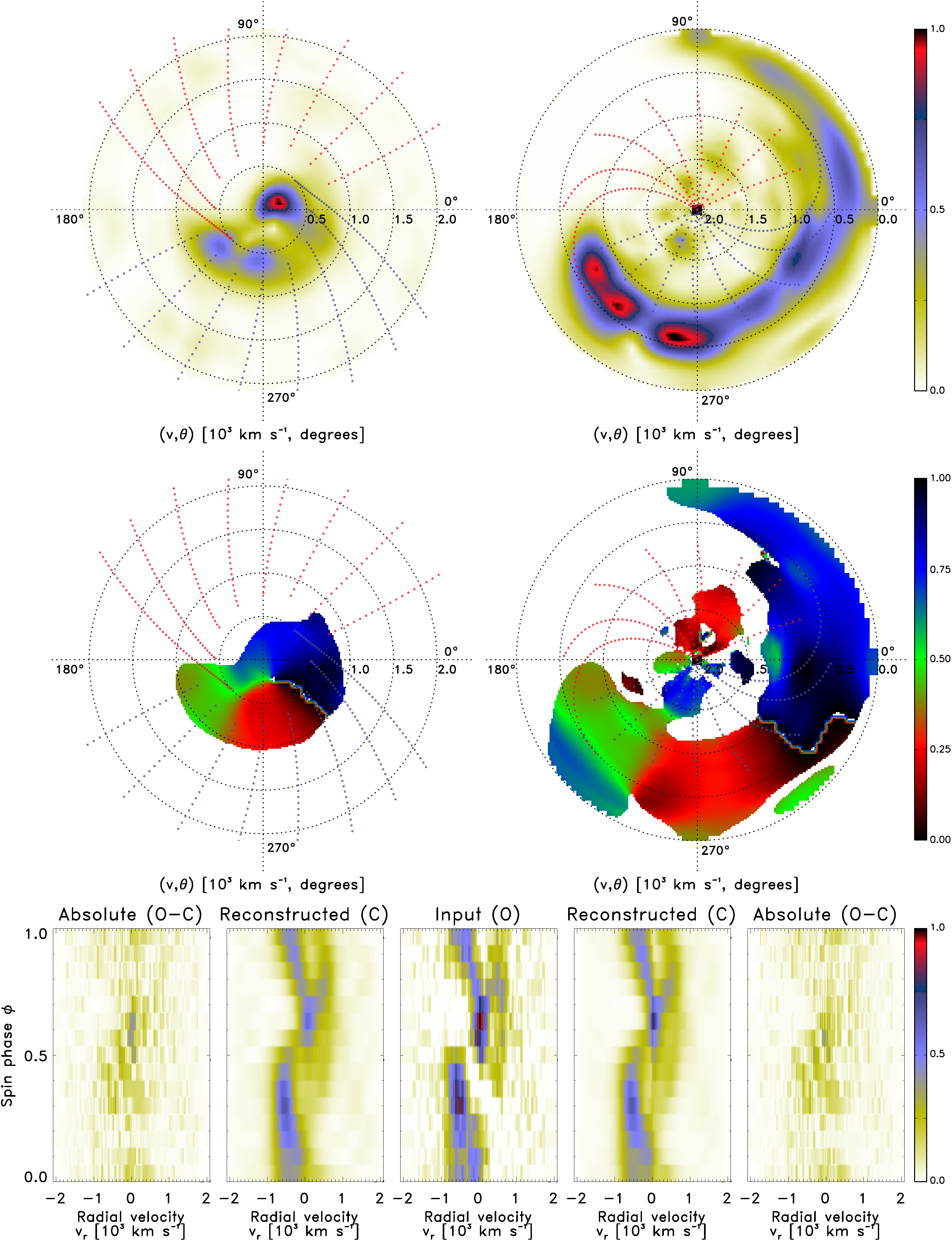}
\caption
 {
  Flux modulation mapping of PQ Gem.
  Same layout as Figs \ref{fig:fmHUAqr} and \ref{fig:fmV834Cen}.
  The model velocity profile overlay shown in all the tomograms is the same as
  the one in Fig.~\ref{fig:dtPQGem}.
 }\label{fig:fmPQGem}
\end{figure*}
Figure \ref{fig:fmPQGem} shows standard and inside-out flux modulation mapping
of PQ Gem.
Compared to the reconstructed spectra in Fig.~\ref{fig:dtPQGem} the summed
reconstructed spectra better reproduce the input spectra.
Even though the flux levels between phases $0.25-0.75$ are not as high as
required, we see that the flux distribution in the two distinct emission
components is significantly improved.
This is especially apparent in the split of the two components between phases
$0.5-1.0$.
The absolute $\mbox{O}-\mbox{C}$ spectra of both projections clearly show the
improvement in the reproduction of the input spectra with rms values of $0.070$
and $0.066$ compared to the non-modulated rms values of $0.094$ and $0.092$.

We see in both the standard and inside-out amplitude maps that most of the
emission components seen in the Doppler tomograms are heavily flux modulated.
In the standard amplitude map the emission associated with the lower accretion
curtain (represented by the spot in the upper right quadrant) appears to be the
most flux modulated component.
The emission associated with the upper accretion curtain (represented by the
spot in the lower left quadrant) is also flux modulated, but it is split into
two modulating components.
In the inside-out amplitude map this emission is also split into two components,
but now they appear to be the most modulated.

In both the standard and inside-out phase of maximum flux maps we see that the
proposed upper curtain is mostly green, which means that it appears brightest at
spin phase $0.5$ when we have a lateral view of the curtain.
In the inside-out map the low ($<500\mbox{\,km\,s}^{-1}$) and high
($>1500\mbox{\,km\,s}^{-1}$) parts of the proposed upper curtain are blue.
This implies that the `start' (as it leaves the orbital plane) and `end' (as
it falls onto the primary) of the curtain appear brightest at spin phase $0.75$
when these parts flow, respectively, towards and away from us.
The proposed lower curtain is mostly blue in both phase of maximum flux maps,
that is to say, it appears brightest at spin phase $0.75$ when the curtain is
predominantly flowing towards us.
In the inside-out map the high velocity ($>1200\mbox{\,km\,s}^{-1}$) part of the
proposed lower curtain (as it falls onto the primary) is red, which means that
it appears brightest at spin phase $0.25$ when this part is predominantly
flowing away from us.
The arc of emission running through the lower right quadrant that `connects' the
proposed upper and lower curtains is mostly black and red, in other words, it
appears brightest around spin phases $0.0-0.25$.

\section{Summary}
\label{sec:Summary}

We applied our inside-out velocity projection to published data of the polars
HU Aqr, V834 Cen and the intermediate polar PQ Gem.
We note that the inside-out projection leads to a redistribution in the relative
contrast levels in and amongst emitting components such as the ballistic and
magnetically confined accretion flows.
The inside-out projection exposes low velocity emission details which are overly
compacted in the standard projection.
Similarly, it enhances high velocity emission details which are washed out in
the standard projection.

In the inside-out tomogram of HU Aqr the ballistic accretion stream appears
discernible to one and a quarter times the velocity extent observed for it in
the standard tomogram.
The blended emission from the secondary and ballistic stream seen in the
standard tomogram of V834 Cen is more separated in the inside-out tomogram.
Also, the separation between the lower velocity emission from the ballistic
accretion stream and that from the magnetically confined accretion is more
pronounced compared to the standard tomogram.
More striking in the inside-out tomogram, however, is the funnelling of the
emission from the magnetically confined accretion flow as it falls towards the
primary.
As expected, the inside-out tomogram of PQ Gem shows an extended projection of
the lower velocity ($<1000\mbox{\,km\,s}^{-1}$) emission associated with the
accretion curtains.
These results, especially for HU Aqr and V834 Cen, show that the inside-out
projection has the ability to reveal emission not discernible in the standard
projection.
Any extra emission information obtained with the inside-out projection has the
potential to help us form a more complete picture of the emission components in
these types of systems.

We also applied our flux modulation mapping technique, in both the standard and
inside-out projections, to the above mentioned test cases.
We found that the information obtained through the amplitude of the phased
modulation and the phase of maximum observed flux neatly confirms the current
expectation of how the emission components in the test cases modulate over an
observed orbital or spin phase.

We conclude that the inside-out projection complements the existing standard
projection.
Together with flux modulation mapping, it forms a useful addition for
unravelling the different emission line components in the observed spectra of
mCVs.
In the spirit of open access, we are making our code publicly accessible. It
can be downloaded from
\begin{center}
 \url{http://www.saao.ac.za/~ejk/doptomog/main.html}
\end{center}

\begin{acknowledgements}
We thank Axel Schwope and Coel Hellier for providing the HU Aqr and PQ Gem
data, respectively.
Also, we thank the referee for the constructive comments.

This material is based upon work supported financially by the National
Research Foundation.
Any opinions, findings and conclusions or recommendations expressed in this
material are those of the author(s) and therefore the NRF does not accept any
liability in regard thereto.
\end{acknowledgements}


\end{document}